\documentclass[a4paper,11pt]{article}
\pdfoutput=1
\usepackage{jcappub}
\usepackage[T1]{fontenc}
\usepackage{xspace}
\usepackage{multirow, makecell}
\usepackage{tablefootnote}
\usepackage{soul}
\graphicspath{{plots/}}

\usepackage[normalem]{ulem}

\newcommand{\nua}[1]{\ensuremath{\rlap{\kern-2.5pt\ensuremath{\overset{\scriptscriptstyle(-)}{\phantom{\nu}}}}{\ensuremath{{\nu}_{#1}}}}\xspace}

\newcommand{\e}[1]{\ensuremath{\times10^{#1}}}

\newcommand{\logA}{\ensuremath{\log(10^{10}A_s)}}

\newcommand{\mlight}{\ensuremath{m_{\rm lightest}}}
\newcommand{\thud}{\ensuremath{\sin^2\theta_{12}}}
\newcommand{\thut}{\ensuremath{\sin^2\theta_{13}}}
\newcommand{\thdt}{\ensuremath{\sin^2\theta_{23}}}
\newcommand{\dmsq}{\ensuremath{\Delta m_{21}^2}}
\newcommand{\dmsqoev}{\ensuremath{\dmsq\: [10^{-5}\eVq]}}
\newcommand{\DMsq}{\ensuremath{|\Delta m_{31}^2|}}
\newcommand{\DMsqoev}{\ensuremath{\DMsq\: [10^{-3}\eVq]}}
\newcommand{\doublebeta}{\ensuremath{0\nu\beta\beta}}
\newcommand{\nme}[1]{\ensuremath{\mathcal{M}^{0\nu}_{#1}}}

\newcommand{\eVq}  {\text{eV}^2}
\newcommand{\lnBsim}[1]{\ensuremath{\ln(B_{\rm NO,IO})\simeq#1}}

\newcommand{\modno}{\ensuremath{\mathcal{M}_{\rm NO}}}
\newcommand{\modio}{\ensuremath{\mathcal{M}_{\rm IO}}}
\newcommand{\Bnoio}{\ensuremath{B_{\rm NO,IO}}}
\newcommand{\Zno}{\ensuremath{Z_{\rm NO}}}
\newcommand{\Zio}{\ensuremath{Z_{\rm IO}}}

\newcommand{\myinput}[1]{
  \IfFileExists{#1}{\input{#1}}{ }}
\newcommand{\nuparamconstraints}[2]{
    & $\input{results/#1/#2/maxmargedist.tex}$
    & $\input{results/#1/#2/limdo68.tex}$--$\input{results/#1/#2/limup68.tex}$%
    & $\input{results/#1/#2/limdo95.tex}$--$\input{results/#1/#2/limup95.tex}$%
    & $\input{results/#1/#2/limdo99.tex}$--$\input{results/#1/#2/limup99.tex}$%
}
\newcommand{\cosmoparamconstraints}[4]{
    & $\input{results/#1/#2/#3.tex}$
    & $\input{results/#1/#2/limdo#4.tex}$ -- $\input{results/#1/#2/limup#4.tex}$
}

\title{Neutrino masses and their ordering: Global Data, Priors and Models}

\author[a]{S.\ Gariazzo,}
\author[b]{M.\ Archidiacono,}
\author[a]{P.F.\ de Salas,}
\author[a]{O.\ Mena,}
\author[a]{C.A.\ Ternes,}
\author[a]{and M.\ T\'ortola}

\affiliation[a]{Instituto de F\'{\i}sica Corpuscular
(CSIC-Universitat de Val\`{e}ncia)\\ 
Parc Cient\'{\i}fic UV, C/ Catedr\'atico Jos\'e Beltr\'an, 2\\ E-46980 Paterna (Valencia), Spain}

\affiliation[b]{Institute for Theoretical Particle Physics and Cosmology (TTK)\\
RWTH Aachen University, D-52056 Aachen, Germany}

\emailAdd{gariazzo@ific.uv.es}
\emailAdd{archidiacono@physik.rwth-aachen.de}
\emailAdd{pabferde@ific.uv.es}
\emailAdd{omena@ific.uv.es}
\emailAdd{chternes@ific.uv.es}
\emailAdd{mariam@ific.uv.es}

\abstract{We present a full Bayesian analysis of the combination of
current neutrino oscillation, neutrinoless double beta decay and
Cosmic Microwave Background observations.
Our major goal is to carefully investigate the possibility to single out one neutrino mass ordering,
namely Normal Ordering 
or Inverted Ordering, with current data.
Two possible parametrizations (three neutrino masses versus the lightest neutrino mass plus the
two oscillation mass splittings) and priors (linear versus logarithmic) are
exhaustively examined.
We find that the preference for NO is only driven by neutrino oscillation data. Moreover, the values of the Bayes factor indicate that the evidence for NO is strong only when the scan is performed over the three neutrino masses with logarithmic priors; for every other combination of parameterization and prior, the preference for NO is only weak.
As a by-product of our Bayesian analyses, we are able to \textit{(a)} compare the Bayesian bounds on the neutrino mixing
parameters to those obtained by means of frequentist approaches, finding a very good agreement; 
\textit{(b)} determine that the lightest neutrino mass plus the two mass splittings parametrization, motivated by the physical observables, is
strongly preferred over the three neutrino mass eigenstates scan
and \textit{(c)} find that logarithmic priors guarantee a weakly-to-moderately more efficient sampling of the parameter space.
These results establish the optimal strategy to
successfully explore the neutrino parameter space, based on the use of
the oscillation mass splittings and a logarithmic prior on the lightest neutrino mass, when combining neutrino oscillation data with cosmology and neutrinoless double beta decay.
We also show that the limits on the total neutrino mass $\sum m_\nu$ can
change dramatically when moving from one prior to the other.
These results have profound implications for future
studies on the neutrino mass ordering,
as they crucially state the need for
self-consistent analyses which
explore the best parametrization and priors,
without combining results that involve different assumptions.
}
  
\begin{document}
\maketitle
\flushbottom

\section{Introduction}
\label{sec:intro}
Neutrinos are light elementary particles which exclusively interact via weak
interactions, and therefore they decouple from the thermal bath
in the early Universe as extremely relativistic states,
constituting a hot dark matter component in our Universe.
Earth-based experiments have demonstrated
that neutrinos oscillate~\cite{McDonald:2016ixn,Kajita:2016cak}
and therefore that neutrinos are massive particles,
implying the first departure from
the Standard Model (SM) of Particle Physics.
Neutrinos, as hot dark matter particles,
possess large thermal velocities~%
\footnote{A rough estimate of the thermal velocity of
  the neutrino relics with $1$~eV masses is $100$~km/s, which is
  comparable to the typical velocity dispersion of a galaxy.
  Dwarf galaxies have much smaller velocity dispersions,
  implying that neutrinos can not cluster to form
  the smaller structures we observe in our Universe.
  Cold dark matter instead has a negligible
  velocity dispersion, contributing to clustering at all
  scales.},
cluster only at scales below their free streaming scale
and consequently do not contribute to structure formation at small 
scales~\cite{Lesgourgues:2006nd,Lesgourgues-Mangano-Miele-Pastor-2013,Lattanzi:2017ubx}.
The Cosmic Microwave Background (CMB)
is also affected by the presence of massive neutrinos,
as these particles may turn non-relativistic around the time of photon decoupling.
Since neutrino oscillation physics is only sensitive to the squared
mass differences ($\Delta m_{ij}^2=m_i^2-m_j^2$), 
we cannot reliably compute the contribution of relic neutrinos
to the (hot) dark matter of the Universe
until we establish the absolute scale of neutrino masses.
Current oscillation data can be remarkably well-fitted in terms of two
squared mass differences, dubbed as the
solar mass splitting
($\Delta m_{21}^2\simeq 7.6\times 10^{-5}$~eV$^2$)
and the atmospheric mass splitting
($|\Delta m_{31}^2|\simeq 2.5\times 10^{-3}$~eV$^2$)~\cite{deSalas:2017kay}. 
Thanks to matter effects in the Sun, we know that $\Delta m_{21}^2>0$.\footnote{Note that 
the observation of matter effects in the Sun constrains the product $\Delta m_{21}^2\cos 2\theta_{12}$ to be positive. Therefore, depending on the convention chosen to describe solar neutrino oscillations, matter effects either fix the sign of the solar mass splitting $\Delta m_{21}^2$ or the octant of the solar angle $\theta_{12}$, with $\Delta m_{21}^2$ positive by definition.}
Since the atmospheric mass splitting $\Delta m_{31}^2$
is currently measured only via neutrino oscillations in vacuum,
which exclusively depend on its absolute value,
its sign is  unknown at the moment.
As a consequence, we have two possibilities for the ordering
of neutrino masses,
which can be
\emph{normal} ($\Delta m_{31}^2>0$)
or \emph{inverted} ($\Delta m_{31}^2<0$).
Future terrestrial experiments aim at measuring the sign of $\Delta m_{31}^2$
exploiting the matter effects in Earth
using long baseline accelerators~\cite{Acciarri:2015uup}
and atmospheric neutrino experiments~\cite{Aartsen:2014oha,Adrian-Martinez:2016fdl}.

The measurement of the neutrino mass ordering, together with the
determination of the amount of CP violation in the lepton sector and
the extraction of the neutrino mass nature (Dirac versus Majorana)
are mandatory steps to unravel the neutrino mass mechanism and
the origin of neutrino masses.
Neutrinoless double beta decay (\doublebeta), the
rarest nuclear weak process in which the lepton number is violated by
two units, could be realised in nature only if neutrinos are
Majorana~\cite{Schechter:1981bd,GomezCadenas:2011it,DellOro:2016tmg}.
A positive detection of the so-called effective Majorana
neutrino mass would not only firmly assess that neutrinos are Majorana particles,
but could also potentially establish the neutrino mass ordering.
This is also related to the fact that
we can put constraints on the absolute scale of neutrino masses.
Depending on the mass ordering,
a lower bound on the sum of the three active neutrino
masses ($\Sigma m_\nu$) is established.
In the Normal Ordering (NO) we have
$\Sigma m_\nu \gtrsim 0.06$~eV,
while in the Inverted Ordering (IO)
$\Sigma m_\nu \gtrsim 0.10$~eV,
where the exact numbers depend on the uncertainties
on the squared mass differences.
Should the experimental measurements on $\Sigma m_\nu$ be strong enough
to rule out the parameter space corresponding to IO,
we would know that the neutrino mass ordering must be normal.
Cosmological studies can also be used to set upper bounds on $\Sigma
m_\nu$ combining CMB data with different large scale structure observations,
providing the bound $\Sigma m_\nu<0.12$~eV at $95\%$~C.L.~\cite{Palanque-Delabrouille:2015pga,DiValentino:2015sam,Cuesta:2015iho,Vagnozzi:2017ovm}.
The possible ways we can exploit to determine the neutrino mass ordering,
therefore, include
\textit{a)} future oscillation facilities;
\textit{b)} future neutrinoless double beta decay facilities,
which could have also the potential to unravel the hierarchical
pattern~\cite{Gerbino:2015ixa,DellOro:2016tmg};
and \textit{c)} next-generation of
CMB and large scale structure surveys, which will notably 
improve the present cosmological limits on $\Sigma m_\nu$~\cite{Archidiacono:2016lnv,Hamann:2012fe} and
consequently strongly constrain the IO case
or even determine the mass ordering.

Last but not least, the current efforts on the development of an experiment
devoted to the direct detection of relic neutrinos,
the ``PonTecorvo%
\footnote{Since the current idea is to move the proposal to the Gran Sasso laboratories in Italy, the original name referring to the previous Princeton location has been recently changed.}
Observatory for Light, Early-universe, Massive-neutrino Yield'' (PTOLEMY) proposal~\cite{Betts:2013uya},
may lead to the determination of the absolute scale, the nature and the ordering of neutrino masses
through a completely different way.
PTOLEMY will use the mechanism of neutrino capture on $\beta$-decaying nuclei~\cite{Cocco:2007za}
in order to detect the small number of events due to the interaction of massive relic neutrinos with
the $\sim100$~g of tritium in the detector.
While the energy position in the electron spectrum of the peak due to relic neutrino scattering
depends on the absolute scale of neutrino masses, the event rate may give information
on the Dirac or Majorana nature of neutrinos,
assuming that we will be able to properly compute the enhancement due to the
clustering of relic neutrinos in the neighbourhood of the Milky Way~\cite{deSalas:2017wtt,Zhang:2017ljh}.
While the expected energy resolution of PTOLEMY
will probably not allow to probe the non-degenerate case and distinguish the three different peaks
in the electron spectrum due to the different neutrino mass eigenstates,
which would provide us clean information on the mass ordering~\cite{Long:2014zva},
the method will open the way for a future experiment to achieve this result.

Recently, plenty of work in the
literature has been devoted to test whether
a preference for one ordering over the other,
given current data, exists.
Namely,
in Refs.~\cite{Hannestad:2016fog,Gerbino:2016ehw,Capozzi:2017ipn}, careful
and dedicated analyses of current cosmological data,
either alone or
combined with oscillation measurements, 
have conservatively found Bayesian odds
favouring normal versus inverted
ordering at the level of $2:1$ in~\cite{Hannestad:2016fog} and $3:2$ in~\cite{Gerbino:2016ehw},
which indicate an extremely weak preference for the
normal scheme.
Also, a modest $2\sigma$ preference was found by the
frequentist approach of Ref.~\cite{Capozzi:2017ipn}.
The authors of
Ref.~\cite{Wang:2017htc} agree with these findings,
as they conclude
that their Bayes factor between the normal and the inverted
neutrino mass scenarios is not large enough to establish a clear
preference for one ordering over the other.

By combining the results
from the oscillation global fits carried out previously
in Ref.~\cite{Esteban:2016qun}
and the cosmological constraint $\Sigma m_\nu<0.13$~eV at $95\%$~C.L.~\cite{Palanque-Delabrouille:2015pga,DiValentino:2015sam,Cuesta:2015iho,Vagnozzi:2017ovm},
adopting a different choice for the parametrization and the priors
than those assumed in
Refs.~\cite{Hannestad:2016fog,Gerbino:2016ehw,Capozzi:2017ipn}, 
the exercise of Ref.~\cite{Simpson:2017qvj}
resulted in a strong evidence for the normal
ordering scenario, with odds $42:1$.%
\footnote{It is interesting to note that the same result,
using our version of the Jeffreys' scale (see table~\ref{tab:jeffreys}),
would be marked as ``moderate''.}
This result was immediately debated in Ref.~\cite{Schwetz:2017fey},
where it was argued that the odds $42:1$
are almost completely determined by the use of the logarithmic prior.

More recently, in Ref.~\cite{Caldwell:2017mqu},
the authors considered neutrinoless double beta decay measurements
\cite{Agostini:2017iyd,Albert:2014awa,KamLAND-Zen:2016pfg}
together with cosmological data \cite{Ade:2015xua} and
oscillation results \cite{Esteban:2016qun}
and they found only very mild preference for the normal ordering,
even if the assumed priors were also logarithmic
(as those of Ref.~\cite{Simpson:2017qvj}).
In this case, however, the authors considered a different parametrization,
based on the ($m_{\rm{lightest}}$, $\Delta m^2_{21}$, $|\Delta m^2_{31}|$) parameters,
which differs from the one addressed in Ref.~\cite{Simpson:2017qvj},
where the authors chose to scan over the three neutrino masses ($m_1$, $m_2$, $m_3$).

The major goal of this study is to fully clarify the issue
of the preference for one of the mass orderings which can be
extracted from the current
neutrino oscillation, cosmological and neutrinoless double beta 
decay measurements,
describing how the assumptions
on linear or logarithmic priors and
on the two possible parametrizations described above
can influence the results.
Following this approach,
we will ensure a self-consistent analysis and, consequently,
self-consistent and reliable results,
avoiding arbitrary assumptions and methods.

The structure of this manuscript is the following.
In section~\ref{sec:method} we describe the method
that we adopt in our calculations:
the Bayesian framework we are working with (\ref{sub:stat})
and
the adopted parametrizations (\ref{sub:params}).
Section~\ref{sec:data} is devoted to describe the experimental
constraints we employ in our analysis, which include
neutrino oscillation (\ref{sub:nudata}),
neutrinoless double beta decay (\ref{sub:0n2b})
and cosmological (\ref{sub:cosmodata}) data.
In section~\ref{sec:results}
we review our results:
the Bayesian constraints on the mixing parameters
compared with the frequentist results (\ref{sub:bayesian_numix}),
the limits on the sum of the neutrino masses (\ref{sub:mnu})
and the perspectives for the mass ordering determination (\ref{sub:ordering}).
In section~\ref{sec:conc} we draw our conclusions.

\section{Method and data}
\label{sec:method}

\subsection{Statistical analysis}
\label{sub:stat}
In the following, we shall briefly summarise the statistical analysis
carried out here, based on a full Bayesian approach.
For an extensive
and illustrative discussion of the application of Bayesian methods in
cosmology,
we refer the reader to the review presented in
Ref.~\cite{Trotta:2008qt}.
Our major aim here is to test the two
possible neutrino mass scenarios by means of
model selection techniques.
In the Bayesian framework,
this translates into the
calculation of the Bayesian evidence $Z$, also called the \emph{marginal likelihood},
defined as the average of the
likelihood $p(d|\theta, \mathcal{M})$
under a prior $p(\theta|\mathcal{M})$,
for a specific model $\mathcal{M}$,
a set of parameters $\theta$
and the dataset $d$~\cite{Trotta:2008qt}:
\begin{equation}\label{eq:bayesevidence}
 Z =
 p(d|\mathcal{M}) =
 \int_{\Omega_\mathcal{M}}
 p(d|\theta, \mathcal{M}) \,
 p(\theta|\mathcal{M})\,
 d\theta~.
\end{equation}
Applying Bayes' theorem, one can obtain the model posterior probability:
\begin{equation}\label{eq:modelpostprob}
 p(\mathcal{M}|d)  \propto p(\mathcal{M})\, p(d|\mathcal{M})~,
\end{equation}
where $p(\mathcal{M})$ is the prior probability associated
to the model under consideration.
In our analysis
we will compare the models corresponding to the
normal and inverted mass orderings,
named \modno\ and \modio,
to which we assign identical prior probabilities
$p(\modno)=p(\mathcal{M}_{\rm IO}) = 0.5$.
When comparing the two given models,
the quantity we are
interested in is the ratio of the posterior probabilities, given by:

\begin{equation}\label{eq:modelcomparison}
 \frac{p(\modno|d)}{p(\modio|d)}
 =
 \Bnoio
 \frac{p(\modno)}{p(\modio)}~,
\end{equation}
where the quantity \Bnoio\ is the well-known \emph{Bayes factor},
i.e.\ the ratio of the evidences of the two models:
\begin{equation}\label{eq:bayesfactor}
 \Bnoio = \frac{\Zno}{\Zio} \quad\Longrightarrow\quad \ln \Bnoio = \ln \Zno -\ln \Zio~.
\end{equation}
If the Bayes factor is larger than $1$ ($\ln \Bnoio>0$),
observations favour \modno\ versus \modio.
If $\Bnoio<1$ ($\ln \Bnoio<0$) instead, data would favour \modio\ versus
\modno.
The strength of the preference for one of the competing models over the
other is usually determined by means of the Jeffreys' scale
(see Ref.~\cite{Trotta:2008qt} and references therein),
shown in table~\ref{tab:jeffreys}. 

\begin{table}
\centering
\begin{tabular}{c|c|c}
\hline
$|\ln \Bnoio|$ & Odds & strength of evidence \\
\hline
$<1.0$         & $\lesssim 3:1$ & inconclusive \\
$\in[1.0,2.5]$ & $(3-12):1$     & weak \\
$\in[2.5,5.0]$ & $(12-150):1$   & moderate \\
$>5.0$         & $>150:1$       & strong \\
\hline
\end{tabular}
\caption{Jeffreys' scale for estimating the strength of the preference
  for one model over the other (from Ref.~\cite{Trotta:2008qt}) when
  performing Bayesian model comparison analysis.}
\label{tab:jeffreys}
\end{table}

In order to compute the Bayesian evidence, we use
the \texttt{PolyChord} nested
sampler~\cite{Handley:2015fda,Handley:2015aa},
which can be integrated in the publicly available
\texttt{CosmoMC} code,
that includes the Boltzmann
equation solver CAMB~\cite{Lewis:1999bs,Lewis:2002ah}
for the computation of cosmological quantities.
These are the tools required to compute the Bayes factor,
which will allow for a 
proper model comparison of the normal versus the inverted neutrino mass ordering.

\subsection{Parametrizations}
\label{sub:params}
In order to derive the joint constraints from neutrino oscillations,
\doublebeta\
and cosmological measurements we explore a vast
parameter space,
which consists of up to sixteen physical parameters.
To obtain robust conclusions we describe the neutrino masses under several different assumptions, as detailed below.
\begin{table}[t]
\centering
\begin{tabular}{c|c||c|c||c|c}
\multicolumn{2}{c||}{Cosmological} & \multicolumn{2}{c||}{\doublebeta{}} & \multicolumn{2}{c}{Neutrino mixing} \\
\hline
Parameter & Prior & Parameter & Prior & Parameter & Prior \\
\hline
$\Omega_bh^2$ & 0.019 -- 0.025 & $\alpha_2$       & 0 -- $2\pi$  & $\thud$ & 0.1 -- 0.6  \\
$\Omega_ch^2$ & 0.095 -- 0.145 & $\alpha_3$       & 0 -- $2\pi$  & $\thut$ & 0.00 -- 0.06\\
$\Theta_s$ & 1.03 -- 1.05   & $\nme{^{76}{\rm Ge}}$  & 4.07 -- 4.87 & $\thdt$ & 0.25 -- 0.75\\
$\tau$     & 0.01 -- 0.4    & $\nme{^{136}{\rm Xe}}$ & 2.74 -- 3.45 \\
$n_s$      & 0.885 -- 1.04 \\
$\logA$    & 2.5 -- 3.7    \\
\end{tabular}
\caption{Cosmological parameters, \doublebeta\ parameters and neutrino mixing angles used in the analysis, with the adopted priors.}
\label{tab:commonParams}
\end{table}

The first set of parameters which we assume all throughout the analysis
are the typical six parameters related to the standard cosmology within a flat $\Lambda$CDM universe:
\begin{equation}\label{parameter}
\{\Omega_{\textrm{b}}h^2,\Omega_{\textrm{c}}h^2, \Theta_s, \tau, n_s, \log[10^{10}A_{s}]\}~.
\end{equation}
They are the
baryon $\Omega_{\textrm{b}}h^2$ and the cold dark matter
$\Omega_{\textrm{c}}h^2$ energy densities,
the ratio between the sound
horizon and the angular diameter distance at decoupling $\Theta_{s}$,
the reionization optical depth $\tau$,
the scalar spectral index
$n_s$ and the amplitude $A_{s}$ of the primordial power spectrum.
The
priors adopted on these six parameters are reported in the left column of
table~\ref{tab:commonParams}. 

Concerning the neutrino mixing parameters, there are three angles and one CP-violating phase $\delta$ in the PMNS
neutrino mixing matrix~\cite{Giunti:2007ry}.
Given that current neutrino oscillation measurements are
unable to set strong constraints on the value of $\delta$, in our simulations
we will ignore this parameter, which moreover does not affect  the cosmological or \doublebeta\  observables.
In any case, the global neutrino oscillation results  adopted here are
computed marginalising away the CP violating phase.
Concerning the three mixing angles, we use the physical
parameters $\thud$, $\thut$ and $\thdt$, which are the ones directly involved
in the three-neutrino framework oscillation probabilities, with the priors
shown in the right column of table~\ref{tab:commonParams}. The prior ranges were chosen to fully cover the $3\sigma$ region around the best fit for both mass orderings.

Current bounds from \doublebeta\ experiments are provided in terms of the so-called
effective Majorana mass of the electron neutrino:
\begin{equation}
m_{\beta\beta} = \frac{m_e}{\mathcal{M^{0\nu}}\sqrt{G_{0\nu}T^{0\nu}_{1/2}}}~,
\label{eq:thalf}
\end{equation}
where $T^{0\nu}_{1/2}$ is the \doublebeta\ half-life,
$m_e$ is the electron mass,
$G_{0\nu}$ is a
phase-space factor and
$\mathcal{M^{0\nu}}$ is the Nuclear Matrix
Element (NME),
a delicate and crucial quantity whose uncertainty can
strongly affect the derived bounds on $m_{\beta\beta}$ from
\doublebeta\ searches.
Here we follow the approach of
Ref.~\cite{Caldwell:2017mqu}, which deals with the results reported
by the Gerda \cite{Agostini:2017iyd},
EXO-200 \cite{Albert:2014awa} and
KamLAND-Zen \cite{KamLAND-Zen:2016pfg} experiments.
Therefore, we have two extra
common parameters in our analyses,
which encode the uncertainty on the NMEs of
$^{76}$Ge (in the case of Gerda)
and $^{136}$Xe (in the case of EXO and KamLAND-Zen).
Following
\emph{The Modest Proposal for the Ranges of Values of the NMEs} provided by the authors of Ref.~\cite{Giuliani:2012zu},
we adopt the range
$[4.07-4.87]$ for $^{76}$Ge
and
$[2.74-3.45]$ for $^{136}$Xe,
as reported in the central column of table~\ref{tab:commonParams}.
We have verified that using extended ranges for the NMEs
does not  affect significantly our results on the neutrino mass ordering.
The other two extra parameters required
for the \doublebeta\ analysis are related to the definition of the effective Majorana mass
\begin{equation}
m_{\beta\beta}
= \left|\sum_{k} e^{i\alpha_k}\,
U_{\rm PMNS,ek}^2\, m_k\right|~,
\label{eq:meff}
\end{equation}
where $\alpha_k$ ($k=1,\,\ldots,\,3$) are the Majorana phases,
which play no role in
neutrino oscillations but are a basic ingredient in \doublebeta\ 
processes.\footnote{%
In the normal ordering case, some combinations of
the Majorana phases could lead to an accidental
cancellation of $m_{\beta\beta}$.}
One of the phases can always be rotated away,
therefore we are left with two Majorana phases,
which we choose to be $\alpha_2$ and $\alpha_3$.
As done in previous related works~\cite{Gerbino:2015ixa,Gerbino:2016ehw},
we apply flat priors in the range $[0,\,2\pi]$ for these two Majorana
phases, whose values are totally unknown,
as also reported in table~\ref{tab:commonParams}.

\begin{table}[t]
\centering
\begin{tabular}{c|c|c||c|c|c}
\multicolumn{3}{c||}{Case A} &
\multicolumn{3}{c}{Case B} \\
\hline
Parameter & Prior & Range & Parameter & Prior & Range \\
\hline\hline
\multirowcell{ 2}{$m_1/\mathrm{eV}$} & linear & 0 -- 1         & \multirowcell{ 2}{\mlight/eV} & linear & 0 -- 1         \\
                            & log    & $10^{-5}$ -- 1 &                               & log    & $10^{-5}$ -- 1 \\
\hline
\multirowcell{ 2}{$m_2/\mathrm{eV}$} & linear & 0 -- 1         & \multirowcell{ 2}{\dmsq/eV$^2$} & \multirowcell{ 2}{linear} & \multirowcell{ 2}{$5\e{-5}$ -- $10^{-4}$} \\
                            & log    & $10^{-5}$ -- 1 &                                 & \\
\hline
\multirowcell{ 2}{$m_3/\mathrm{eV}$} & linear & 0 -- 1         & \multirowcell{ 2}{\DMsq/eV$^2$} & \multirowcell{ 2}{linear} & \multirowcell{ 2}{$1.5\e{-3}$ -- $3.5\e{-3}$} \\
                            & log    & $10^{-5}$ -- 1 &                                 & \\
\hline
\end{tabular}
\caption{
Parametrizations of the neutrino masses and priors adopted in the analysis.
}
\label{tab:massParams}
\end{table}

The latest parameters we need to account for are related to the
description of neutrino masses.
Even if the underlying theoretical model is the same,
the physical parameter space can be described in different ways,
which can be mapped one into the other using some (non-linear) transformations.
While the physics does not change, different parameterizations can have different performances when comparing the theoretical model
with the experimental data,
which reflects in different Bayesian evidences.
Here we distinguish two
possible approaches, that we label as
\textit{Case A} and \textit{Case B}.
For the former one, Case A,
we perform the scan over the individual neutrino masses
(i.e.\ $m_1$, $m_2$ and $m_3$),
following the approach of Refs.~\cite{Gerbino:2016ehw,Simpson:2017qvj}.
The latter one, Case B,
focuses on the ($m_{\rm{lightest}}$, $\Delta m^2_{21}$, $\Delta m^2_{31}$) parameter
space~\cite{Caldwell:2017mqu}.
Being physically equivalent, the choice of one or the other parameterization
basically reflects in different structures of the parameter space and of the prior function.
Comparing Case A and Case B, therefore, can be seen as
comparing the efficiency of different priors,
non-linearly connected one to the other,
in describing the available data.
For both parametrizations we study linear and logarithmic priors
on the physical mass parameters,
in order to take into account that the true absolute mass scale is unknown,
while we always use a linear prior
for the squared mass differences.
The complete list of priors is reported in 
table~\ref{tab:massParams}.
Additionally, we will explore how variations in the prior ranges
affect the final results.
The comparison of the Bayesian evidences
obtained with different priors and parametrizations
will allow us to avoid results which are biased
by subjective arbitrary choices in the analyses.

As we shall show in the following sections, given the current available
measurements, the
Bayesian analyses show that data prefer the informative priors of
Case B over the general priors of Case A.
In Case B, the priors
are chosen to match the available information from neutrino oscillation experiments on the measured mass splittings
and they ensure a very efficient scan in the parameter space.%
\footnote{The situation would be different without the information coming from the neutrino oscillation experiments, which severely constrain the relation between the three individual masses.}
The situation is very similar to the one that we need to
face when exploring the mixing angles:
it is more efficient to scan the parameter space using
the $\sin^2\theta_{ij}$ quantities than using the mixing angles
$\theta_{ij}$ themselves.
While very futuristic galaxy and 21cm surveys might be able to
disentangle the individual values of the neutrino
masses~\cite{Pritchard:2008wy},
present cosmological measurements are only sensitive
to the total neutrino mass and
the information on the three masses can be extracted
only using the input from neutrino oscillation experiments.
Since in Case A the parameter space is described using
the three masses $m_i$,
it is less efficiently explored than the parameter space
of Case B, which includes the two squared mass differences.

\section{Experimental data}
\label{sec:data}

\subsection{Neutrino oscillation data}
\label{sub:nudata}

\begin{table}[t]
\centering
{
\renewcommand{\arraystretch}{1.2}
\catcode`?=\active \def?{\hphantom{0}}
\begin{minipage}{\linewidth}
\begin{tabular}{l||c|c|c}
parameter & best fit $\pm$ $1\sigma$ &  2$\sigma$ range& 3$\sigma$ range
\\
\hline\hline
$\Delta m^2_{21}\: [10^{-5}\eVq]$ & 7.56$\pm$0.19  & 7.20--7.95 & 7.05--8.14 \\
\hline
$|\Delta m^2_{31}|\: [10^{-3}\eVq]$ (NO) &  2.55$\pm$0.04 &  2.47--2.63 &  2.43--2.67\\
$|\Delta m^2_{31}|\: [10^{-3}\eVq]$ (IO)&  2.47$^{+0.04}_{-0.05}$ &  2.39--2.55 &  2.34--2.59 \\
\hline
$\sin^2\theta_{12} / 10^{-1}$ & 3.21$^{+0.18}_{-0.16}$ & 2.89--3.59 & 2.73--3.79\\
\hline
  $\sin^2\theta_{23} / 10^{-1}$ (NO)
	  &	4.30$^{+0.20}_{-0.18}$ \footnote{There is a local minimum in the second octant, at 
	    $\sin^2\theta_{23}$=0.596 with $\Delta\chi^2 = 2.08$ with respect to the
	    global minimum.} 
	& 3.98--4.78 \& 5.60--6.17 & 3.84--6.35 \\
  $\sin^2\theta_{23} / 10^{-1}$ (IO)
	  & 5.98$^{+0.17}_{-0.15}$ \footnote{There is a local minimum in the first octant, at 
	    $\sin^2\theta_{23}$=0.426 with $\Delta\chi^2 = 3.0$ with respect to the
	    global minimum for IO.} 
	& 4.09--4.42 \& 5.61--6.27 & 3.89--4.88 \& 5.22--6.41 \\
\hline 
$\sin^2\theta_{13} / 10^{-2}$ (NO) & 2.155$^{+0.090}_{-0.075}$ &  1.98--2.31 & 1.89--2.39 \\
$\sin^2\theta_{13} / 10^{-2}$ (IO) & 2.155$^{+0.076}_{-0.092}$ & 1.98--2.31 & 1.90--2.39 \\
    \hline
  \end{tabular}
  \caption{ \label{tab:sum-2017} 
    Neutrino oscillation parameters summary determined from the 
    global analysis in~\cite{deSalas:2017kay}. The ranges for inverted ordering refer to the 
    local minimum within this neutrino mass ordering.}
    \end{minipage}
  }
\end{table}

\begin{figure}[t]
\centering
\includegraphics[width=0.95\textwidth]{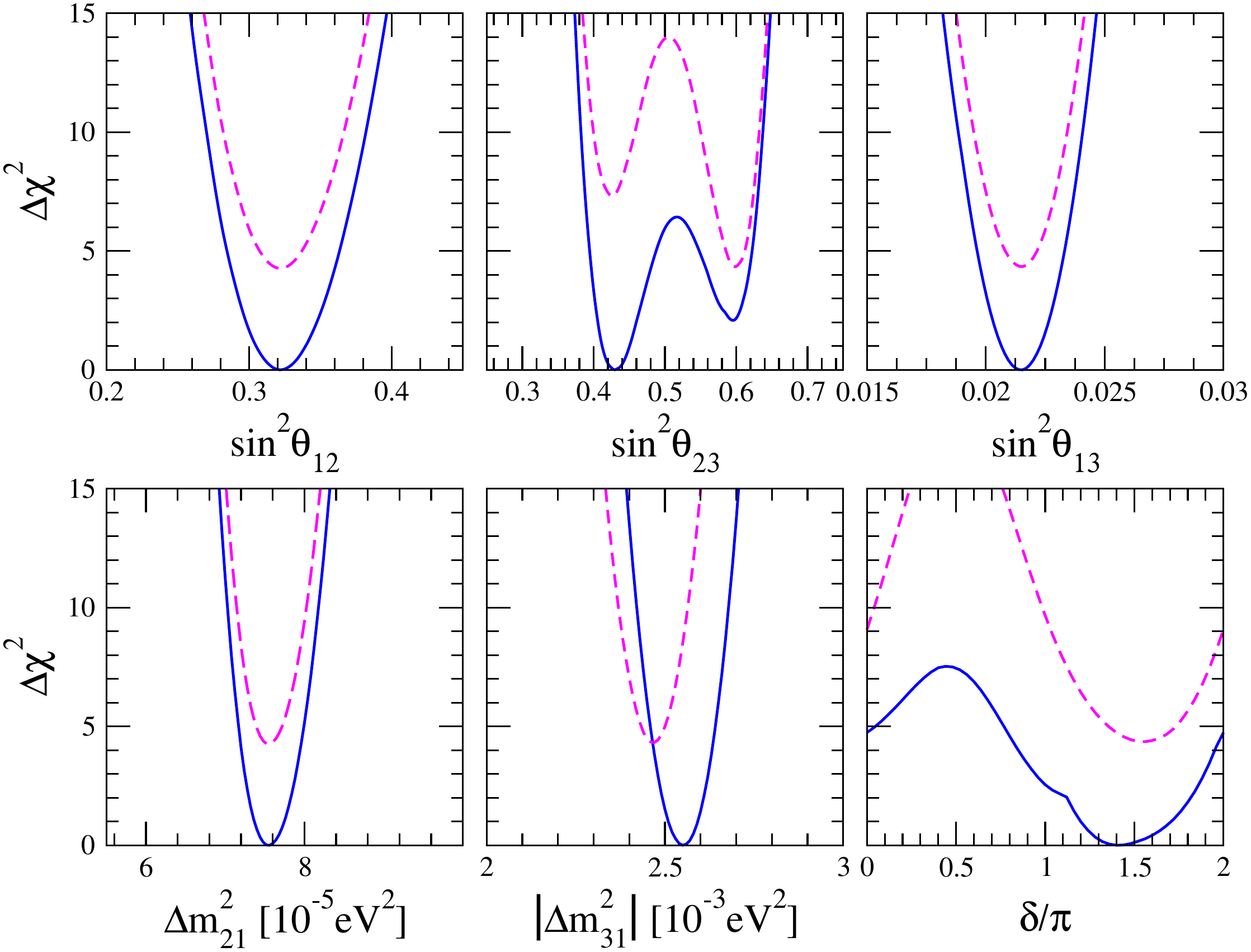}
\caption{The profiles for the neutrino oscillation parameters. Solid
blue lines correspond to NO and  dashed magenta lines to IO.}
\label{fig:panel-dchi2}
\end{figure}

Our combined analysis is based on the 
global fit of neutrino oscillation parameters performed in Ref.~\cite{deSalas:2017kay}.
For our Bayesian calculations, we adopt the $\chi^2_{\rm osc}$ as a function of the mixing parameters
and squared mass differences and we convert it into a likelihood $\mathcal{L}_{\rm osc}$ using:
\begin{equation}\label{eq:oscllh}
\log\mathcal{L}_{\rm osc} = -\chi^2_{\rm osc}/2\,.
\end{equation}
Figure~\ref{fig:panel-dchi2} and table~\ref{tab:sum-2017}
summarise the results of the global analysis~\cite{deSalas:2017kay}.
To perform the global fit we have analysed data from all  neutrino oscillation experiments.  
The parameters  $\Delta m_{21}^2$ and $\sin^2\theta_{12}$ are measured with great precision by the solar 
experiments~\cite{Cleveland:1998nv,Kaether:2010ag,Abdurashitov:2009tn,Hosaka:2005um,Cravens:2008aa,Abe:2010hy,Nakano:PhD,Aharmim:2008kc,Aharmim:2009gd,Bellini:2013lnn}
and the long baseline reactor experiment KamLAND~\cite{Gando:2010aa}.
The reactor neutrino data that we use comprise 
the 1230 days of data taken by Daya Bay~\cite{An:2016ses}, 500 live days of RENO data~\cite{RENO:2015ksa} and the Double Chooz  far detector data from the far detector only--period (461 days) and the  far and near detector--period (212 days)~\cite{Abe:2014bwa}. 
These short--baseline reactor experiments are mainly sensitive to $\sin^2\theta_{13}$, whose determination in the global fit is dominated by
the measurements of  Daya Bay. Moreover, the most recent reactor results  show a good sensitivity to the atmospheric mass splitting $\Delta m_{31}^2$ as well. 
The global fit to neutrino oscillations includes also atmospheric data from the 
phases I to III of the Super-Kamiokande experiment~\cite{Wendell:2010md} and the 
neutrino telescopes IceCube DeepCore~\cite{Aartsen:2014yll,IC-www} and ANTARES~\cite{AdrianMartinez:2012ph}. Note, however, that the data and detector information needed to reproduce the last IceCube DeepCore results  presented in~\cite{Aartsen:2017nmd} have not been made public yet, so those are not considered in our analysis. We do  not include the atmospheric neutrino results of 
Super Kamiokande phase IV~\cite{Abe:2017aap} either, since there is not enough public information to reproduce the Super-Kamiokande analysis of atmospheric data outside the Collaboration.
As it is explained in Ref.~\cite{deSalas:2017kay}, neutrino telescopes start being competitive with long baseline experiments, although these still perform better when determining neutrino oscillation parameters.
From long baseline accelerator neutrino experiments we include data taken by the already finished K2K~\cite{Ahn:2006zza} and MINOS 
experiments~\cite{Adamson:2013whj,Adamson:2014vgd}, as well as from their successors, T2K~\cite{Abe:2017bay,Abe:2017uxa} and NO$\nu$A~\cite{Adamson:2017qqn,Adamson:2017gxd}.
Besides constraining the atmospheric parameters $\Delta m_{31}^2$ and $\sin^2 \theta_{23}$, 
the long baseline experiments are also sensitive to $\sin^2 \theta_{13}$ and $\delta$.
Indeed, most of the current sensitivity to the CP-violating phase
$\delta$ comes from  the combined analysis of T2K  neutrino and antineutrino data. 
The current best fit value for this parameter tends towards maximal CP-violation, with $\delta=3\pi/2$. 
However, it remains the worst determined neutrino oscillation parameter, with most of the parameter space allowed at the 3$\sigma$ level.
In any case, and as it was mentioned in section~\ref{sub:params}, this parameter is not crucial for the present analysis, so it has been marginalised over. 

Another oscillation parameter that still has room for ambiguities is the atmospheric angle $\theta_{23}$, whose octant 
cannot yet be resolved at the $3\sigma$ level. While T2K data prefer maximal mixing 
$(\sin^2\theta_{23}=0.5)$, the MINOS and NO$\nu$A experiments show a preference for $\theta_{23}$  in the lower octant, resulting in a global best fit value of 
$\sin^2\theta_{23}=0.43$. A local minimum is however present in the upper octant, with maximal mixing disfavoured at more than $2\sigma$ 
confidence level, as can be seen in figure~\ref{fig:panel-dchi2}. 
When inverted mass ordering is assumed, a similar scenario is obtained, 
although in this occasion the global minimum lies in the upper octant and a local minimum for $\theta_{23}$ appears in the lower octant.

Comparing the two mass orderings with  flavour oscillation data only, we obtain a slight preference for normal 
ordering with $\Delta\chi^2=4.3$. If we only consider the long baseline experiments T2K and NO$\nu$A, the difference between normal and inverted ordering is found to be $\Delta\chi^2=3.6$. This result is due to the better agreement between  the T2K and NO$\nu$A preferred values of $\sin^2\theta_{23}$ in normal ordering  with respect to inverted ordering. Adding  the information from atmospheric  data to the fit relaxes the tension between the preferred values of  $\theta_{23}$ by T2K and NO$\nu$A  in IO, to $\Delta\chi^2=3.1$.
The best fit value for $\sin^2\theta_{13}$ from the 
combined analysis of  long baseline  plus atmospheric data lies quite close to global best fit value, mainly fixed by the Daya Bay reactor experiment. However, in the case of normal ordering these two values are much closer to each other than for the inverted ordering case.
Therefore, the inclusion of reactor data in the global neutrino fit increases again the preference for normal  over inverted ordering
to the final value  of $\Delta\chi^2=4.3$. Further details on the determination of  the mass ordering or the atmospheric angle octant from the global fit can be found in Ref.~\cite{deSalas:2017kay}.

\subsection{Neutrinoless double-beta decay data}
\label{sub:0n2b}
We parameterize the constraints from \doublebeta\ searches
following the same approach of Ref.~\cite{Caldwell:2017mqu}.
Therefore, we use a combined likelihood
\begin{equation}\label{eq:doublebetalike}
\mathcal{L}_{\doublebeta}
=
\prod_i
\mathcal{L}_{\doublebeta, i}
\,,
\end{equation}
where the terms $\mathcal{L}_{\doublebeta, i}$
describe the likelihood for each experiment
(Gerda \cite{Agostini:2017iyd},
KamLAND-Zen \cite{KamLAND-Zen:2016pfg} and
EXO-200 \cite{Albert:2014awa}):
\begin{eqnarray}
\mathcal{L}_{\doublebeta,\ \rm{GERDA}}(T_{1/2}^{\rm Ge})
&\propto&
\exp\left[
-\frac{(1/T_{1/2}^{\rm Ge}+1.48)^2}{2\times 0.461^2}
\right]
\,,\label{eq:gerda}\\
\mathcal{L}_{\doublebeta,\ \rm{KamLAND-Zen\ phase\ I}}
(T_{1/2}^{\rm Xe})
&\propto&
\exp\left[
-\left(2.3/T_{1/2}^{\rm Xe}+1.09/(T_{1/2}^{\rm Xe})^2\right)
\right]
\,,\label{eq:kamland1}\\
\mathcal{L}_{\doublebeta,\ \rm{KamLAND-Zen\ phase\ II}}
(T_{1/2}^{\rm Xe})
&\propto&
\exp\left[
-\left(9.71/T_{1/2}^{\rm Xe}+28.1/(T_{1/2}^{\rm Xe})^2\right)
\right]
\,,\label{eq:kamland2}\\
\mathcal{L}_{\doublebeta,\ \rm{EXO}}(T_{1/2}^{\rm Xe})
&\propto&
\exp\left[
-\frac{(1/T_{1/2}^{\rm Xe}-0.32)^2}{2\times 0.30^2}
\right]
\,.\label{eq:exo}
\end{eqnarray}
Here, the half-life decay times are given in units of $10^{25}$ years
and all the parametrizations have been confirmed by
the experimental collaborations~\cite{Caldwell:2017mqu}.
The conversion between the half-life time and the neutrino masses
is written in eqs.~\eqref{eq:thalf} and \eqref{eq:meff}.

\subsection{Cosmological data}
\label{sub:cosmodata}
Planck satellite measurements of the  CMB temperature,
polarization, and cross-correlation spectra from the 2015 release~\cite{Adam:2015rua,Ade:2015xua}
have been included.%
\footnote{We make use of the publicly
  available Planck likelihoods~\cite{Aghanim:2015xee}, see
  \href{http://www.cosmos.esa.int/web/planck/pla}{www.cosmos.esa.int/web/planck/pla}.}
More precisely, we exploit both
the high-$\ell$ ($30 \leq \ell \leq 2508$) $TT$ and
the low-$\ell$ ($2 \leq \ell \leq 29$) $TT$
likelihoods,
based on the reconstructed CMB maps.
The Planck
polarization likelihood in the low-multipole regime
($2 \leq \ell \leq 29$) is added
to the previous two CMB datasets as well.
Furthermore, we also include here the high-multipole
($30 \leq \ell \leq 1996$) $EE$ and $TE$ likelihoods. 
All these likelihood functions have a non-negligible dependence on a
given number of  nuisance parameters
(e.g.\ residual foreground contamination, calibration, and
beam-leakage~\cite{Ade:2015xua,Aghanim:2015xee}),
which have also been
taken into account and properly marginalised over.

\section{Results}
\label{sec:results}

\subsection{Global Bayesian constraints on the neutrino mixing parameters}
\label{sub:bayesian_numix}

\begin{table}[t]
\centering
\resizebox{1\textwidth}{!}{
\renewcommand{\arraystretch}{1.1}
\begin{tabular}{c||c|c|c|c}
parameter & bestfit &
	$1\sigma$ range &
	$2\sigma$ range & $3\sigma$ range \\
\hline\hline
$\dmsqoev$      \nuparamconstraints{B_no/B_no_osc}{deltam2_21} \\
\hline
$\DMsqoev$ (NO) \nuparamconstraints{B_no/B_no_osc}{deltam2_31} \\
$\DMsqoev$ (IO) \nuparamconstraints{B_io/B_io_osc}{deltam2_31} \\
\hline
$\thud$         \nuparamconstraints{B_no/B_no_osc}{theta12} \\
\hline
$\thdt$ (NO)
    & 0.43 (0.59) 
    & 0.40--0.47 \& 0.58--0.60
    & 0.38--0.50 \& 0.53--0.63
    & 0.37--0.65
    \\
$\thdt$ (IO) 
    & 0.60 (0.43) 
    & 0.56--0.63
    & 0.39--0.47 \& 0.54--0.65
    & 0.37--0.66
    \\
\hline
$100\thut$ (NO) \nuparamconstraints{B_no/B_no_osc}{theta13} \\
$100\thut$ (IO) \nuparamconstraints{B_io/B_io_osc}{theta13} \\
\hline
\end{tabular}
}
\caption{
Bayesian results on the neutrino mixing parameters from the global fit of neutrino oscillation data.
The results are in agreement with the corresponding frequentist global fit,
see table~\ref{tab:sum-2017}, taken from ref.~\cite{deSalas:2017kay}. The parameters in parenthesis represent
the $\thdt$ local minima.
}
\label{tab:bayesnufit}
\end{table}

\begin{figure}[t]
 \centering
\includegraphics[width=0.95\textwidth]{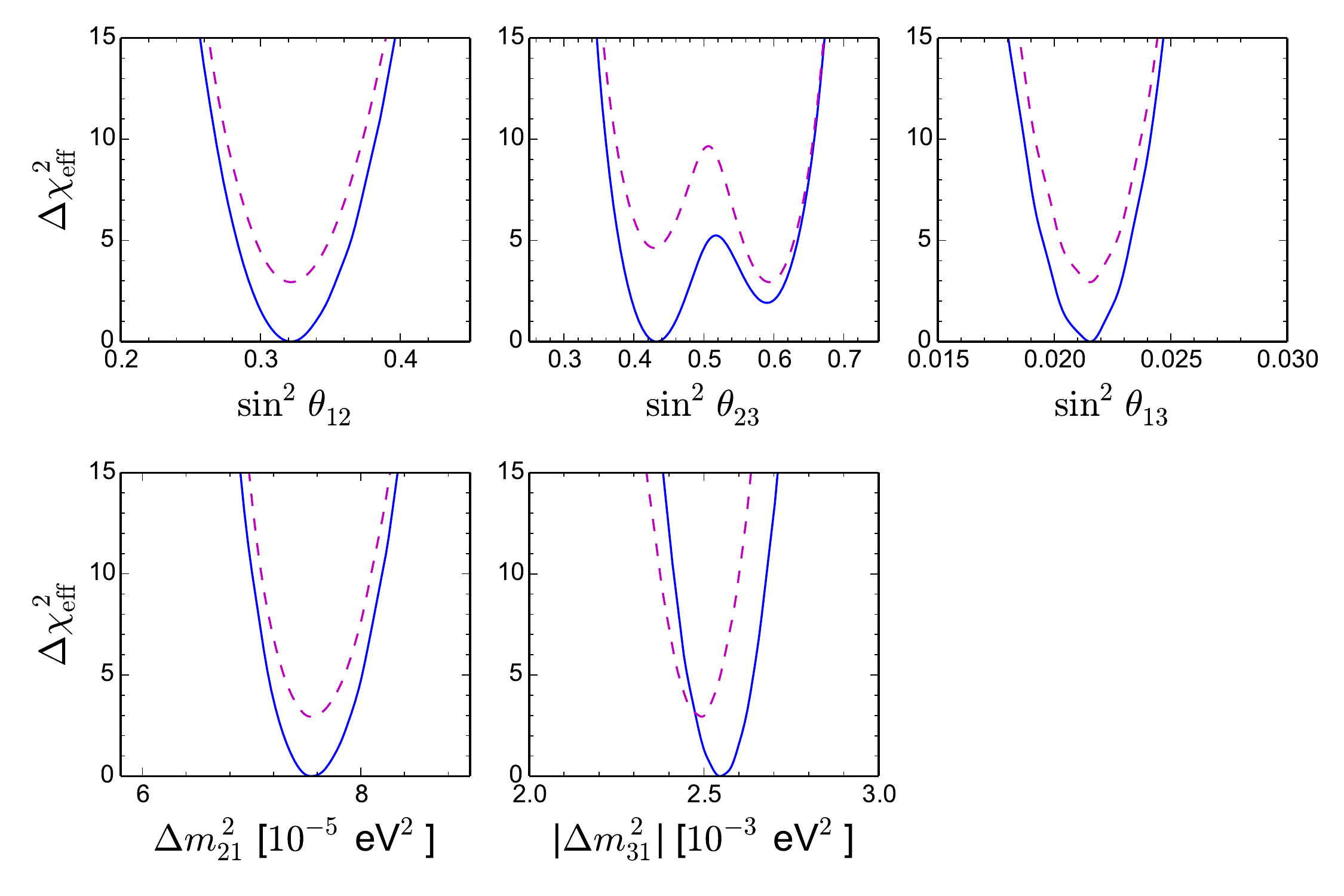}
\caption{The profiles for the neutrino oscillation parameters as obtained from the Bayesian analysis.
Blue (solid) lines correspond to NO and magenta (dashed) lines to IO.
The effective $\chi^2$ is obtained from the posterior distribution function $p(x)$ for each parameter $x$ and the Bayesian evidence $Z$
using
$\Delta \chi^2_{\rm eff} = -2 \ln p(x) -2\ln Z$.}
\label{fig:bayesian1d}
\end{figure}

Our model comparison analysis will provide, as a by-product, a Bayesian
analysis of the neutrino oscillation parameters. These Bayesian limits 
can be compared with the frequentist analysis of Ref.~\cite{deSalas:2017kay}. 
Our results are summarised in table~\ref{tab:bayesnufit} and in figure~\ref{fig:bayesian1d}.
Notice that at the $3\sigma$ level there is no significant
difference with respect to the frequentist method (summarised in table~\ref{tab:sum-2017} and figure~\ref{fig:panel-dchi2}),
except for the atmospheric angle $\thdt$.
In the case of the frequentist approach, maximal mixing is excluded at $3\sigma$ for the case of inverted ordering, 
while it remains allowed in the Bayesian analysis. In addition, the width of the confidence intervals
for $\thdt$ is larger in the Bayesian case than in the frequentist
one for the two mass orderings.
Note also that,  in the Bayesian case, the local minimum found at $\thdt=0.59$ for normal ordering now falls
inside the 1$\sigma$ allowed range, contrary to what happens for inverted ordering or in the frequentist case for both orderings.
To understand the differences between both analyses one should keep in mind that
the marginalisation procedure in Bayesian parameter estimation
is different from the one in the frequentist analysis. In the
Bayesian case, the volume of the posterior distribution is also an
important quantity, together with the relative difference in terms of $\chi^2$ (or,
equivalently, in terms of the likelihood). Therefore, when performing Bayesian marginalisation,
the posterior volume near the two minima (global and local) is large enough to prevent
a strong exclusion of maximal atmospheric mixing in both mass orderings.
On the other hand, since the local minimum  in $\thdt$ is closer to the global one in the NO case ($\Delta\chi^2=2.08$)
than in the IO one ($\Delta\chi^2=3.0$), it still appears inside the 1$\sigma$ region for NO.

For the remaining  mixing parameters, the allowed Bayesian intervals are in very good agreement with the frequentist ones,
with only small deviations which are not significant. 

\subsection{Limits on the sum of the neutrino masses}
\label{sub:mnu}

\begin{table}[t]
\centering
\resizebox{1\textwidth}{!}{
\begin{tabular}{c||c|c||cc||cc}
\multirowcell{ 2}{Dataset} &
\multirowcell{ 2}{Case} &
\multirowcell{ 2}{Prior type} &
\multicolumn{2}{c||}{NO} &
\multicolumn{2}{c}{IO} \\
& & & bestfit & $2\sigma$ range & bestfit & $2\sigma$ range \\
\hline\hline
\multirowcell{ 2}{OSC\\+ \doublebeta}
& \multirowcell{ 2}{B}
  & linear
    \cosmoparamconstraints{B_no/B_no_lin_osc_0n2b}{mnu}{maxmargedist}{95}
    \cosmoparamconstraints{B_io/B_io_lin_osc_0n2b}{mnu}{maxmargedist}{95}
    \\
& & log
    \cosmoparamconstraints{B_no/B_no_log_osc_0n2b}{mnu}{maxmargedist}{95}
    \cosmoparamconstraints{B_io/B_io_log_osc_0n2b}{mnu}{maxmargedist}{95}
    \\
\hline\hline
\multirowcell{ 2}{OSC\\+CMB} 
& \multirowcell{2}{B}
  & linear
    \cosmoparamconstraints{B_no/B_no_lin_osc_cmb}{mnu}{maxmargedist}{95}
    \cosmoparamconstraints{B_io/B_io_lin_osc_cmb}{mnu}{maxmargedist}{95}
    \\
& & log
    \cosmoparamconstraints{B_no/B_no_log_osc_cmb}{mnu}{maxmargedist}{95}
    \cosmoparamconstraints{B_io/B_io_log_osc_cmb}{mnu}{maxmargedist}{95}
    \\
\hline\hline
\multirowcell{ 2}{ALL}
& \multirowcell{ 2}{B}
  & linear
    \cosmoparamconstraints{B_no/B_no_lin_osc_cmb_0n2b}{mnu}{maxmargedist}{95}
    \cosmoparamconstraints{B_io/B_io_lin_osc_cmb_0n2b}{mnu}{maxmargedist}{95}
    \\
& & log
    \cosmoparamconstraints{B_no/B_no_log_osc_cmb_0n2b}{mnu}{maxmargedist}{95}
    \cosmoparamconstraints{B_io/B_io_log_osc_cmb_0n2b}{mnu}{maxmargedist}{95}
    \\
\hline
\end{tabular}}
\caption{\label{tab:mnubounds}
Best fit values and marginalised $2\sigma$ bounds for the sum of the neutrino masses (in eV),
for different datasets, using Case B and the priors shown in table~\ref{tab:massParams}.
The case ALL corresponds to the OSC + CMB + \doublebeta\ combination.}
\end{table}

Currently, the tightest constraints on the sum of the neutrino masses
are provided by cosmological observations. In order to derive
these bounds, the $\Lambda$CDM is typically assumed as the fiducial
model.
In the standard analysis (see
e.g.~\cite{Ade:2015xua,Palanque-Delabrouille:2015pga,
DiValentino:2015sam,Cuesta:2015iho,
Giusarma:2016phn,Vagnozzi:2017ovm}),
the parameter space is usually scanned with a linear prior
on the (degenerate) neutrino masses through their sum $\Sigma m_\nu$.
In this section we shall show that the upper limits on $\Sigma m_\nu$
strongly depend on the prior choice.
Therefore, special attention should be payed when using existing external results, computed by means of a given prior assumption,
to derive constraints within analyses based on a different prior choice.

Our constraints on $\Sigma m_\nu$ are summarised in table~\ref{tab:mnubounds},
where we show the best fit and the marginalised 2$\sigma$ limits for NO and IO, for
different combinations of data sets, for the Case B and for
different choices of priors, as detailed in
section~\ref{sub:params}.
Notice that we have not considered Case A since,
as we will show below,
it is much less efficient in exploring the parameter space with respect to Case B
when considering oscillation data only.
A quick 
inspection of table~\ref{tab:mnubounds} tells us that the
difference in the upper limits obtained
when considering the same prior and data
within normal and inverted
neutrino mass ordering is not very large.%
\footnote{The only exception to this statement, appearing when using the linear prior and OSC+CMB+\doublebeta\ data,
is related to a statistical fluctuation that we found in the \texttt{PolyChord} output.}
This was expected, as
these two cases are equivalent in the degenerate neutrino mass region.

Let us now focus on the crucial role of the prior choice (linear or logarithmic) which,
according to table~\ref{tab:mnubounds},
has a critical
impact on the total
neutrino mass bounds.
Sampling a variable with uniform probability in $\log x$ corresponds to
assuming a probability $1/x$ for $x$ itself,
which leads to a bias in the
scanning towards small values of $x$.
As a consequence, the posterior of $\Sigma m_\nu$ is driven towards the
smallest values allowed by oscillation data when the prior is
logarithmic,
because of the different available volume in the neutrino mass parameter space.
Indeed, the upper bounds in the linear case are very close to those obtained previously by
e.g.\ the Planck collaboration~\cite{Ade:2015xua},
while in the logarithmic case they are much smaller.
Similar conclusions about the prior dependence of the results
have been studied in Ref.~\cite{Hannestad:2017ypp}, where it was shown 
that the proper way to treat constrained parameters
(e.g.\ those forced to be non-negative by physical reasons)
is to employ a Jeffreys prior, which is the one that maximises the sensitivity of the posterior distribution to data.
The exact functional shape of the Jeffreys prior must be derived
using the response of the likelihood to changes in the parameters.%
\footnote{A proper calculation of the Jeffreys
prior should be performed separately for each of the parameterizations and priors we are
adopting, a computation that is beyond the scope of this work.}
More recently, and based on the Lagrangian of
the underlying particle physics theory (i.e.\ on the
neutrino mass matrix), the theoretical priors on $\sum m_\nu$ have
been extracted~\cite{Long:2017dru}. However, we do not implement none of the above prescriptions in computing our
results, as \textit{(a)} $\Sigma m_\nu$ is not a free parameter in our
MCMC analyses, \textit{(b)} the neutrino oscillation likelihood is
multivariate in \thdt, and
\textit{(c)} individual runs for the Dirac, Majorana and Majorana
seesaw cases, which are beyond our main purposes, are mandatory while following the prescription of
Ref.~\cite{Long:2017dru}. 

A final comment is devoted to the studies in the literature which employ the CMB posterior distributions obtained in the
standard $\Lambda$CDM$+\Sigma m_\nu$ model
as a prior for other analyses.
Let us firstly comment that the problem of the prior affecting
the $\Sigma m_\nu$ bounds is a consequence of the fact
that we adopt the mass of the lightest neutrino,
which is unconstrained from below, as a parameter in our analyses.
For standard cosmological analyses,
since neutrino oscillations do constrain $\Sigma m_\nu$ from below,
it is not necessary to consider a logarithmic prior on $\Sigma m_\nu$
and a linear one is sufficient,
even though it may not be the more appropriate
one~\cite{Hannestad:2017ypp} (see above).
As demonstrated in table~\ref{tab:mnubounds}, the CMB is not powerful enough to constrain $\Sigma m_\nu$~\cite{Archidiacono:2016lnv}
and therefore the posterior on this parameter depends heavily on the
prior.
Indeed, in Bayesian analysis, for a given data set, the posterior is the convolution of the prior with the likelihood. If the likelihood is sufficiently informative, the posterior has a very weak dependence on the prior. In the opposite case, data cannot provide sufficient information and the analysis returns something similar to what we knew before, i.e.\ the prior.
Consequently, it is extremely dangerous to
use a posterior distribution obtained previously with specific priors
when performing an analysis that is based on different ones, 
because the internal consistency is lost.
The situation might change in the future:
if new data will have the power to constrain the mass of the lightest neutrino from below,
the effect of the prior assumptions will vanish.

\subsection{Implications for the mass ordering}
\label{sub:ordering}
\begin{figure}[tp]
\centering
\includegraphics[width=0.99\textwidth]{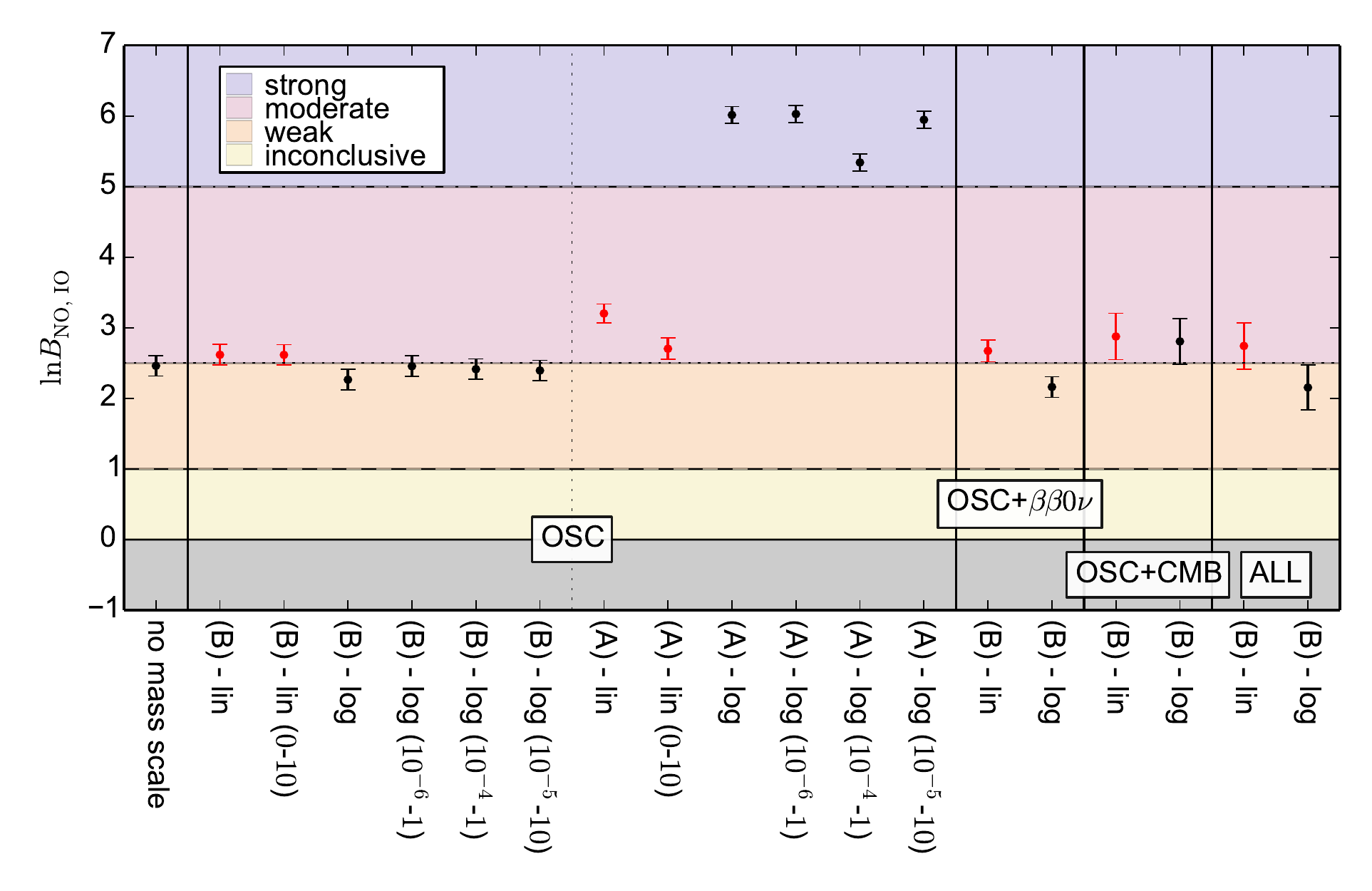}
\caption{\label{fig:bayesfactors}
Graphical visualisation of the Bayesian factors comparing normal and inverted ordering.
The horizontal lines indicate the values at which there is a change in the statistical significance
of the evidence, according to the Jeffreys' scale (see table~\ref{tab:jeffreys}).
Black (red) points indicate a logarithmic (linear) prior. The prior ranges are those reported in table~\ref{tab:massParams} if not otherwise stated.}
\end{figure}

We shall focus in the following on the preference
for one neutrino mass ordering over the other from observations. Figure~\ref{fig:bayesfactors} illustrates the Bayes factors for
NO over IO for different prior choices.
First of all, we notice that the preference towards NO
is solely driven by neutrino oscillation data,
regardless of the selected priors.
Indeed, when one simply considers the Bayesian analysis of
the neutrino oscillation data and varies only the five mixing
parameters (first point on the left, named as \emph{no mass scale}),
the preference $\lnBsim{2.5}$ comes from the difference in the
minimum $\Delta\chi^2$ from the global fit of neutrino oscillation
measurements. In the Jeffreys' scale (table~\ref{tab:jeffreys}), this corresponds to a weak-to-moderate preference for NO.

\begin{figure}[tp]
\centering
\includegraphics[width=1.\textwidth]{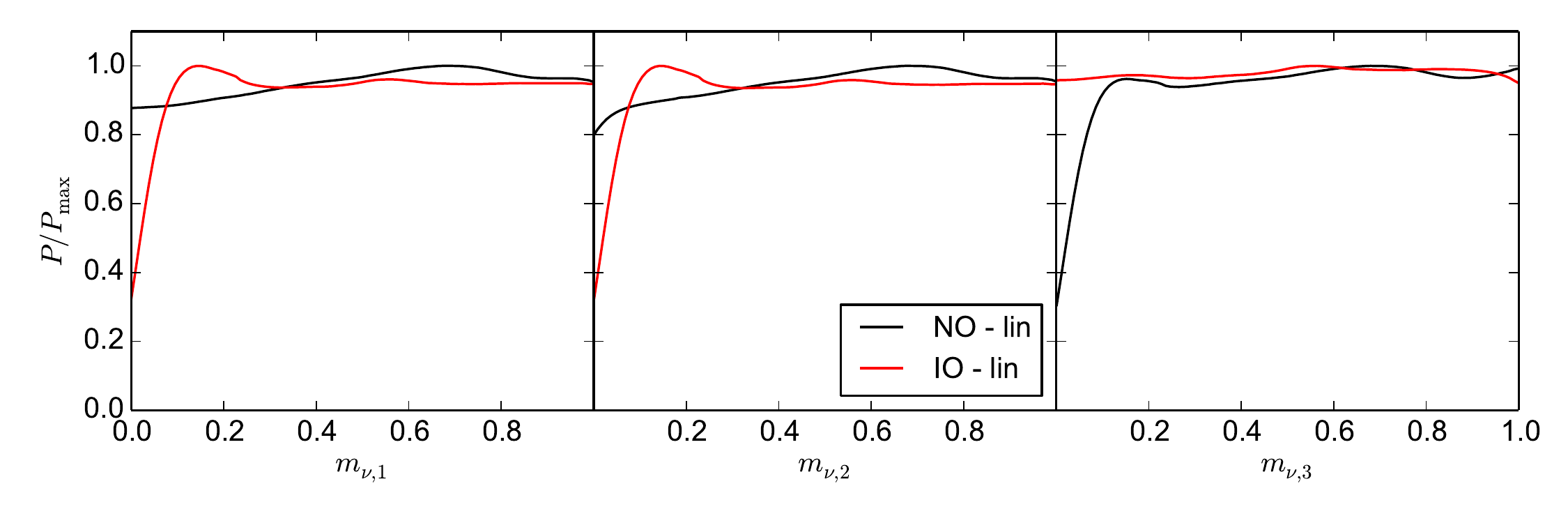}
\includegraphics[width=1.\textwidth]{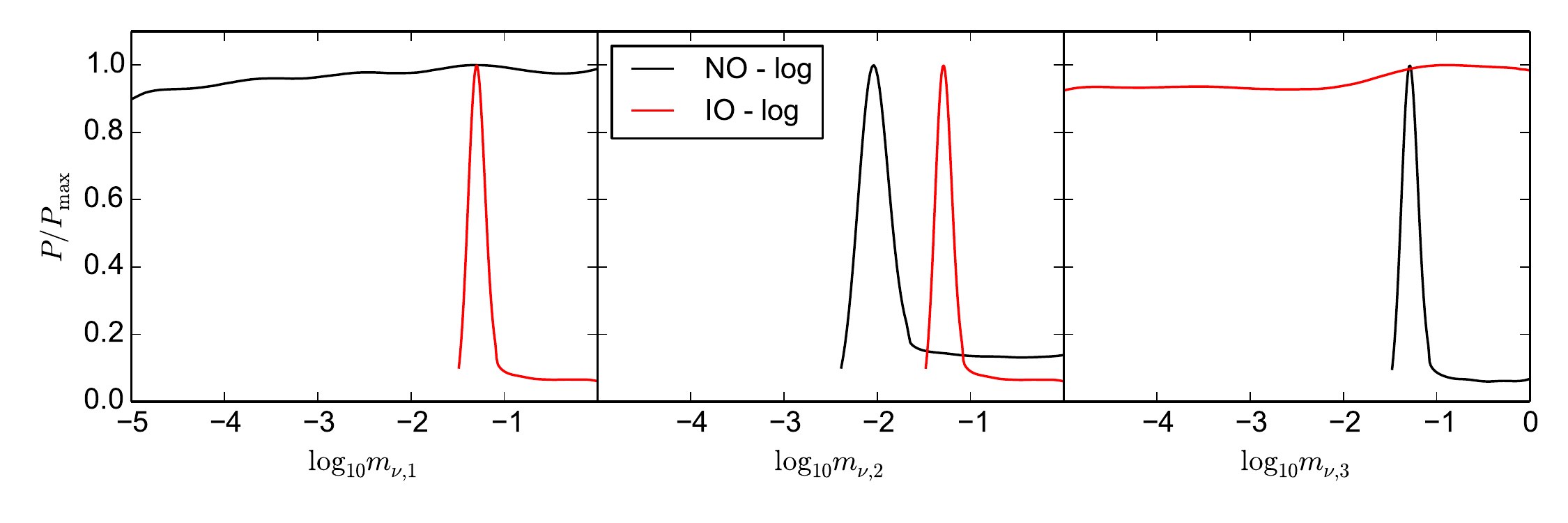}
\caption{\label{img:massesvolumes}
Difference in allowed volumes for the three absolute neutrino masses
for NO and IO from neutrino oscillation data only. The top (bottom) panels show the case of
linear (logarithmic) priors.}
\end{figure}

Focusing on the first block of figure~\ref{fig:bayesfactors}, which
deals with neutrino oscillation data exclusively, one can 
notice that the choice of parameterization and prior has a direct effect on the strength of the
preference of NO versus IO.
While in the case of Case B the Bayes factors are highly stable
against changes in the priors or in the allowed range for the lightest
neutrino mass, for Case A the prior choice can strongly affect
the Bayes factor in favour of NO. This happens because, within
Case A, the volume of the parameter space that corresponds to the
allowed values of the neutrino masses by oscillation data differs
substantially from the total volume, sampled by means of
the three individual neutrino masses, which are allowed to vary in a
vast range. Within Case B this difference does not exist, because the
parameter space is scanned using different variables,
i.e.\ using informative priors. Since the
difference between NO and IO mostly appears at small masses,
prior choices that give more importance to small neutrino masses
(i.e.\ logarithmic ones) will correspond to a stronger preference for
NO within Case A. 
Figure~\ref{img:massesvolumes} shows the posterior distribution for
the three neutrino masses from neutrino oscillation data only, comparing NO
and IO with linear (top panels) and logarithmic (bottom
panels) priors.
In the case of linear priors, the changes induced in the posterior
volumes for the three mass eigenstates are nearly equivalent for NO
and IO, and therefore the Bayes factor will be very close to that extracted from
oscillation data alone. However, as logarithmic priors naturally
increase the importance  of smaller masses,
the difference in the posterior volumes for
NO and IO is much more evident in this case, as can be seen in the lower panels of
figure~\ref{img:massesvolumes}.
The difference in the
behaviour of $m_1$ and $m_3$ is practically irrelevant, as they reverse their roles
in NO and IO. However, the change in the volume due to $m_2$ is crucial:
Indeed, in the case of NO $m_2$ is bounded from below by $\dmsq$,
while in IO the lower bound is given by $\DMsq$.
For this reason, the posterior volume in Case A strongly disfavours
IO when a logarithmic prior is considered, and therefore Case A with logarithmic priors is the only one
in which we find a strong preference for NO.

When we also account for
information on the neutrino mass scale, either from CMB or \doublebeta\
probes, the situation for Case B does not change dramatically
with respect to the oscillations-only case, as shown in
figure~\ref{fig:bayesfactors}.
Case A is not considered because it is
much less efficient than Case B (see below).

\begin{figure}[tp]
\centering
\includegraphics[width=0.99\textwidth]{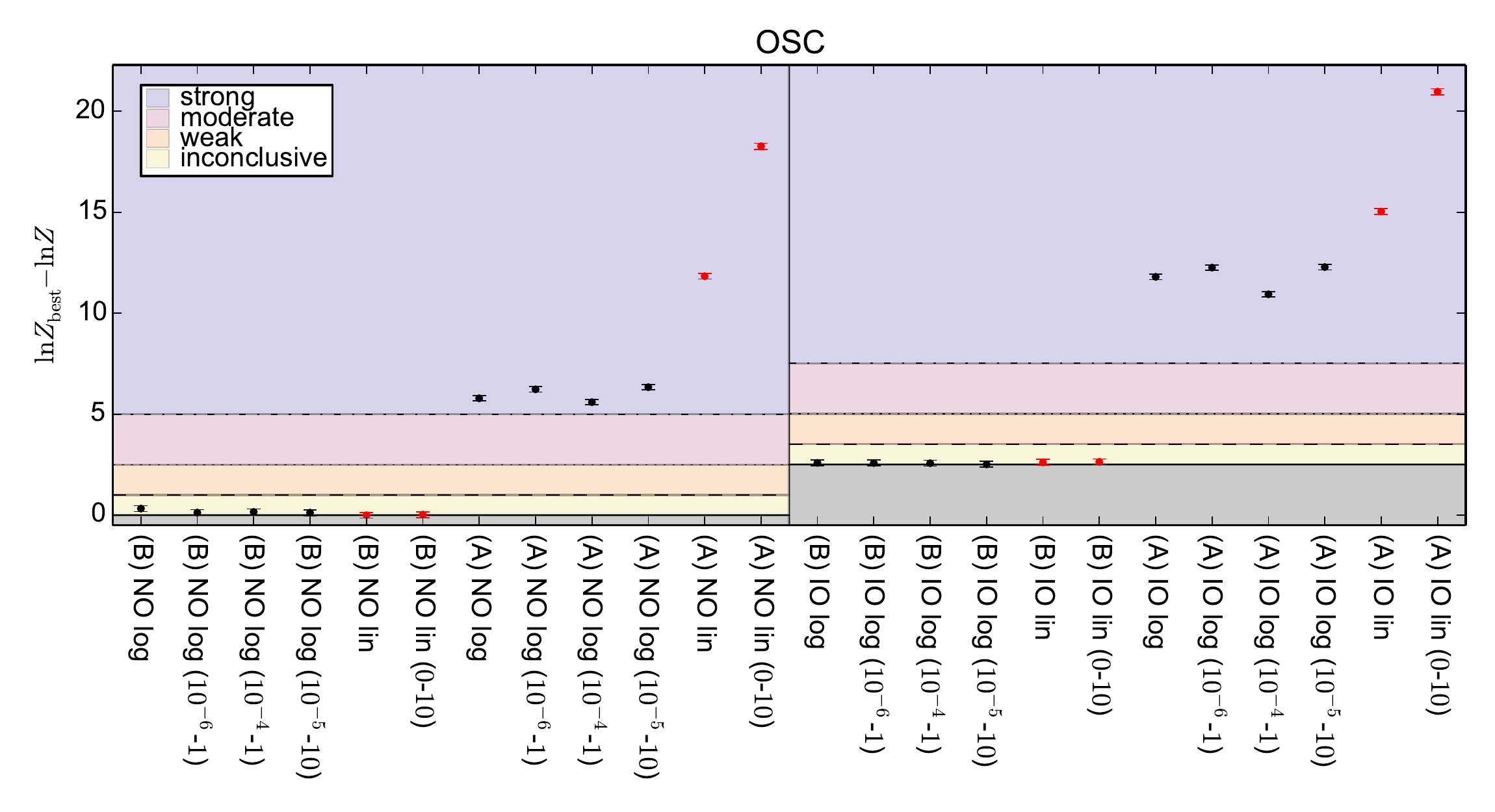}

\includegraphics[width=0.32\textwidth]{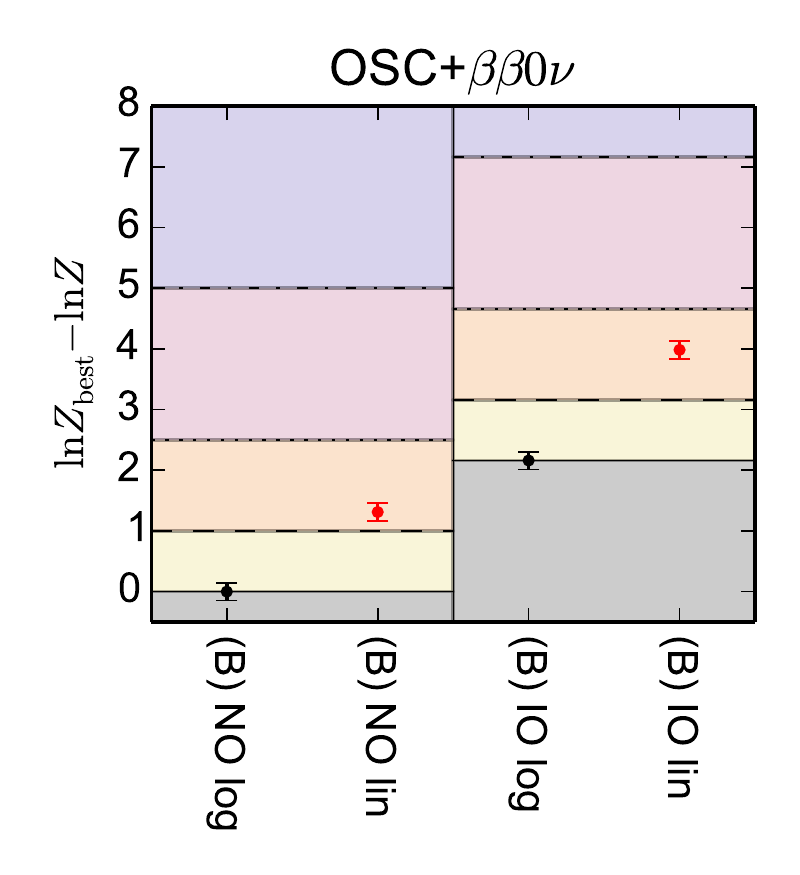}
\includegraphics[width=0.32\textwidth]{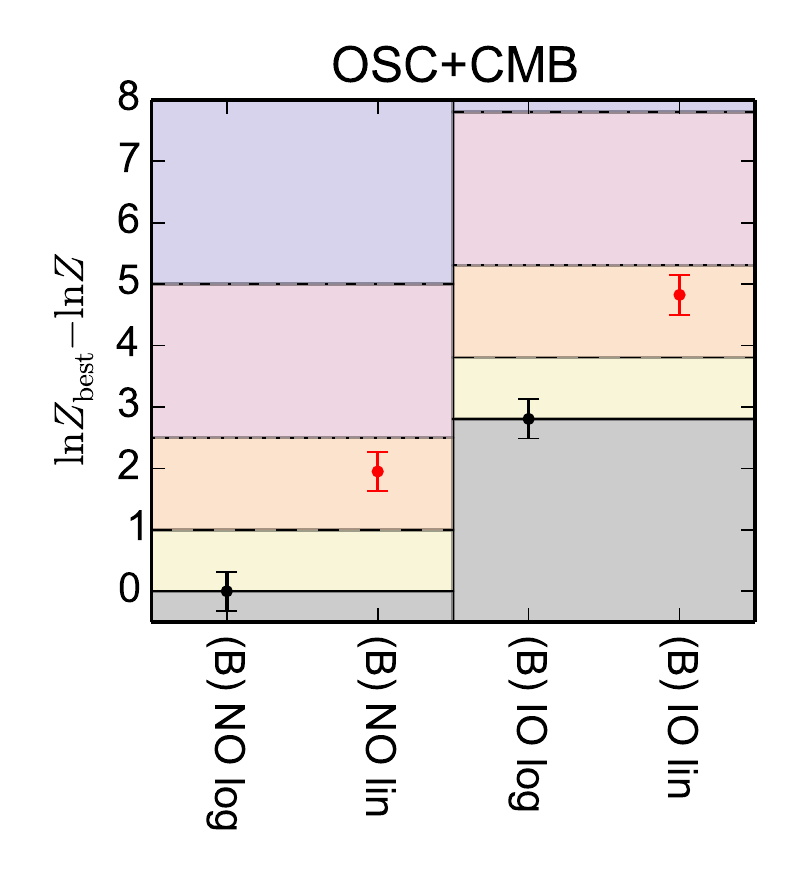}
\includegraphics[width=0.32\textwidth]{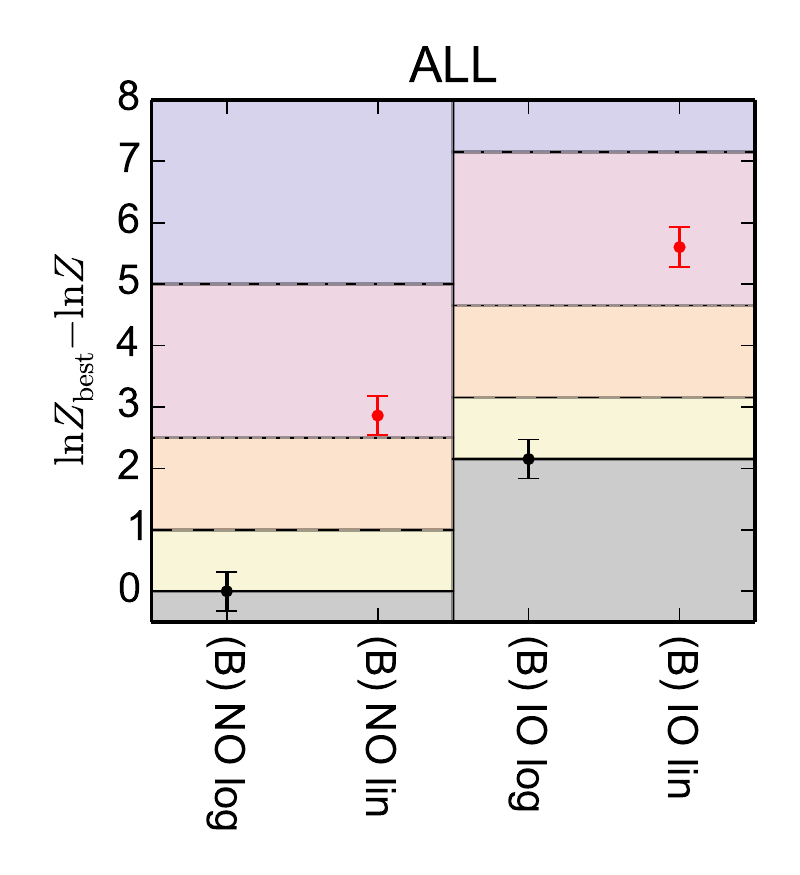}

\caption{\label{fig:bayEvid}
Graphical visualisation of the Bayesian evidences to compare the
various neutrino mass parametrizations (Case A, only
shown in the neutrino oscillation analyses, and Case B,
shown for all the data combinations) and
priors (logarithmic, in black, and linear, in red).
Points are normalised with respect to
the preferred Bayesian evidence $Z_{\rm best}$
within each panel,
which always corresponds to one of the Case B cases with NO.
Colour codes are the same as in figure~\ref{fig:bayesfactors}.}
\end{figure}

The Bayesian evidence analysis also opens the possibility to test 
the various parametrizations (i.e.\ Case A versus Case B) and prior choices (logarithmic versus linear). 
Figure~\ref{fig:bayEvid} illustrates the Bayesian evidence of the different parameterizations and priors,
normalised to the best scenario within each plot, for an easier comparison.
Each of the panels corresponds to a given combination of data sets 
as indicated above and
it is divided in two sub-panels, one for NO (left) and one for IO (right).

The top panel shows how a change in the prior ranges impacts the
Bayesian evidences of Cases A and B, when either a linear or
logarithmic prior is adopted and only oscillation measurements are
considered in the numerical analysis. Notice that a different
prior on the lightest neutrino mass leaves the Bayesian
evidences of Case B unchanged both for NO and IO and for both linear
and logarithmic priors. This is because the
parameter \mlight\ remains unconstrained
in all the cases, and the Bayes factor does not penalise
unconstrained parameters through the Occam's razor.
On the other hand, 
when considering Case A, the linear priors are always
moderately-to-strongly less efficient for the parameter space exploration with respect to logarithmic priors,
and the Case A itself is always much less efficient with respect to Case B.
The reason is simple: the parametrization that uses three neutrino masses
as free parameters corresponds to a waste of parameter space. 
Since neutrino oscillations determine the squared mass differences
with a high accuracy, most of the parameter space in Case A at high
neutrino masses is useless for the data fit, so that this parametrization is indeed penalised by the Occam's razor.
Being motivated by the physical observables, Case B is therefore preferred over Case A when performing the analyses.

The bottom panels of figure~\ref{fig:bayEvid}, which are restricted to Case
B, tell us that the addition of \doublebeta\ or cosmological data introduce a difference in the Bayesian evidences between linear and logarithmic priors.
These data indeed show that the logarithmic priors are weakly-to-moderately more efficient,
because in this latter case the fraction of volume corresponding to small masses,
preferred by the data,
is larger than in the linear case.

\section{Conclusions}
\label{sec:conc}
Plenty of work has been recently devoted in the literature to
infer the neutrino mass ordering using a number of present
observations~\cite{Capozzi:2017ipn, Esteban:2016qun,deSalas:2017kay,
Simpson:2017qvj,
Vagnozzi:2017ovm,Gerbino:2016ehw,Schwetz:2017fey,Hannestad:2016fog,Wang:2017htc,Caldwell:2017mqu},
but a complete and self-consistent Bayesian analysis
was still missing.
Such an analysis is necessary in order to avoid strong claims in favour of normal mass
ordering, based exclusively on the choice of priors.
We have presented here the results obtained from the computationally expensive Bayesian evidence calculations, using current neutrino oscillation data, $\doublebeta$ decay searches and
Cosmic Microwave Background cosmological observations.
In order to
explicitly show the crucial role played by both the prior choice, we analyse two possible parametrizations:
\textit{(a)} Case A, in which the scan is performed over the individual neutrino masses
($m_1$, $m_2$, $m_3$), and \textit{(b)} Case B, which is focused on
the ($m_{\rm{lightest}}$, $\Delta m^2_{21}$, $\Delta m^2_{31}$)
parameter space. For both parametrizations we study linear and logarithmic priors
on the physical mass parameters, while we always use a linear prior
for the squared mass differences.

Focusing first on our main goal, the Bayesian evidence against IO recently claimed~\cite{Simpson:2017qvj} and extensively
debated in Ref.~\cite{Schwetz:2017fey}, we find that the value of the Bayes factor is
$\lnBsim{2.5}$ for almost all the possible parametrizations and prior
choices. This value, which only points to weak evidence for NO, is entirely due to neutrino oscillation data.
There is, however, one \emph{single} combination in which we find strong
evidence for NO, corresponding to the particular case on
which Ref.~\cite{Simpson:2017qvj} was focused on,
i.e.\ when the scan is performed
over the ($m_1$, $m_2$, $m_3$) parameter space with logarithmic
priors. This strong preference for NO arises from the
changes in the volume of $m_2$ between the two mass orderings,
as this parameter is limited from below 
by $\dmsq$ in NO but by $\DMsq$ in IO.
While our cosmological data sets are based on CMB measurements
carried out  by the Planck mission, we do not expect that our main
conclusions change significantly  when adding other possible
observations, as Baryon Acoustic Oscillation (BAO) data.
For instance,
Ref.~\cite{Gerbino:2016ehw} quotes a modest 1.5:1 for the odds
in favour of NO
when BAO data is also considered in the analyses,
corresponding to an uninformative $\ln \Bnoio\simeq0.4$.
Consequently,
our conclusions would be practically unchanged
when adding the BAO information.

Another target of our study was the comparison between the Bayesian bounds on the neutrino mixing
parameters to those obtained by means of frequentist approaches,
finding, in general, a very good agreement.

We have also shown that
oscillation data strongly prefer Case B versus
Case A, as a scan over the ($m_{\rm{lightest}}$, $\Delta m^2_{21}$, $\Delta m^2_{31}$)
parameter space, being motivated by the physical observables,
is more efficient than an analysis based on the three neutrino mass eigenstates.
Furthermore, we have also computed the Bayesian evidence for
logarithmic versus linear priors from current data, reporting
that logarithmic priors guarantee a weakly-to-moderately
more efficient scan of the parameter space.

The results derived here should be regarded as a guidance for future
studies focused on the Bayesian comparison of the two neutrino mass
orderings
within the context of a global analysis involving neutrino oscillation, cosmological and neutrinoless double-beta decay data.
Such studies should use the parametrization involving
the neutrino squared mass splittings plus the lightest neutrino mass with a logarithmic prior.
In addition, future studies should avoid using
pre-computed limits on the total neutrino mass $\sum m_\nu$,
since these can change dramatically when moving from one prior choice
to another.
The most general message of this paper is that Bayesian
model comparison provides us with techniques that can be used in order to
produce self-consistent and unbiased results. However, dedicated and careful analyses exploring all the different priors
and parametrizations are always required in order to ensure robust and reliable conclusions.

\acknowledgments
We thank J.~Lesgourgues and S.~Pastor for very useful suggestions and improvements on the
manuscript and to M.~Hirsch and M.~Sorel for fruitful discussions on neutrinoless double beta decay
and the related nuclear matrix elements.
Work supported by the Spanish grants FPA2014-57816-P, FPA2014-58183-P,
FPA2017-85216-P, FPA2015-68783-REDT, FPA2017-85985-P and SEV-2014-0398
(MINECO), FPU13/03729 (MECD) and PROMETEOII/2014/050, PROMETEOII/2014/084 and GV2016-142 (Generalitat
Valenciana).
OM is also supported by the European
Union's Horizon 2020 research and innovation program under the Marie
Sk\l odowska-Curie grant agreements No.\ 690575 and 674896.

\bibliography{bibliography}

\end{document}